\documentclass[12pt]{article}
      \usepackage{amssymb}
      \usepackage{amsmath}
      \usepackage{epsfig}
      \def\href#1#2{#2}
      \def\IP{\relax{\rm I\kern-.18em P}}
      \newcommand{\beq}{\begin{equation}}
      \newcommand{\eeq}{\end{equation}}
      \newcommand{\beqa}{\begin{eqnarray}}
      \newcommand{\eeqa}{\end{eqnarray}}
      \def\tQ{{\hat Q}} 
      \def\bj{{\bar \jmath}} 
      \def\cA{{\cal A}}
      
      \def\cH{{\cal H}}
      \def\cJ{{\cal J}} 
      \def\mod{{\rm mod}}

      \def\QC{\mathbb{C}}

      \def\QR{\mathbb{R}}
      
      \def\QZ{\mathbb{Z}}
      
      \def\S{{\cal S}} 
      \def\orb{{\mbox{\rm \scriptsize orb}}}
      \def\even{{\mbox{\rm \scriptsize even}}}
      \def\be{ \begin{equation}}\def\ee{ \end{equation}}
      \def\ba{ \begin{eqnarray}}\def\ea{ \end{eqnarray}}
      
      \def\nn{\nonumber}   
      
       \def\Z{\mathbb{Z}}

      \def\cedille#1{\setbox0=\hbox{#1}\ifdim\ht0=1ex \accent'30 #1%
       \else{\ooalign{\hidewidth\char'30\hidewidth\crcr\unbox0}}\fi}
      \def\gaw{Gaw\cedille edzki}

      \def\ew{\hspace*{-1mm}}   \def\ppe{\hspace*{-2.5mm}}
      
      \newcommand{\note}[1]{\raisebox{1ex}{{\footnotesize \sf #1}}}
      \newcommand{\rnote}[1]{\raisebox{1ex}{{\hspace*{-3mm} \scriptsize\sf#1}}
         \hspace*{-4mm}}
      \newcommand{\Fus}[6]{F_{{\scriptstyle #1},{\scriptstyle #2}}
        \hspace*{.3mm}\displaystyle{[} \ew \begin{array}{ll} {\scriptstyle #3 }
        \ppe & {\scriptstyle #4} \ppe \\[-2mm] {\scriptstyle #5}\ppe &
        {\scriptstyle #6}\ew \end{array}\displaystyle{]}}
      
      \newcommand{\Br}[6]{B_{{\scriptstyle #1}{\scriptstyle #2}}
        \hspace*{.3mm}\displaystyle{[} \ew \begin{array}{ll} {\scriptstyle #3 }
        \ppe & {\scriptstyle #4} \ppe \\[-2mm] {\scriptstyle #5}\ppe &
        {\scriptstyle #6}\ew \end{array}\displaystyle{]}}
      
      \renewcommand{\vert}[3]{\displaystyle{(} \begin{array}{rcl}
       \ppe & {\scriptstyle #1} \ppe & \ppe \\[-2mm] \ew {\scriptstyle #2} 
        \ppe &  \ppe & {\scriptstyle
       #3}\ew \end{array} \displaystyle{)}}

      \def\min{{\mbox{\rm min\/}}}
      \def\cH{{\cal H}}

      \def\id{{\rm id}}
      
      \def\cS{{\cal S}}

      \title{\bf Open Strings in Simple Current Orbifolds}
      \author{{\sc Keizo Matsubara$,$\rnote{1} } 
        {\sc  Volker Schomerus$\,$\rnote{2} } \\[2mm]
        {\sc Mikael Smedb\"ack$\,$ \rnote{1}}\\[11mm]  
      \note{1} Institutionen f\"or teoretisk fysik, Uppsala Universitet  
      \\ Box 803, S--75108 Uppsala, Sweden
      \\[2mm]
      \note{2} Max-Planck-Institut f\"ur Gravitationsphysik, 
        Albert-Einstein-Institut 
      \\ Am M\"uhlenberg 1, D--14424 Potsdam, Germany\\[2mm]}
      \date{August 15, 2001}
\begin{document}
      \baselineskip=17pt
      \begin{titlepage} \maketitle  \thispagestyle{empty}

      \vskip5mm
      \begin{abstract}
      We study branes and open strings in a large class of orbifold
      backgrounds using microscopic techniques of boundary 
      conformal field theory. In particular, we obtain factorizing 
      operator product expansions of open string vertex operators for 
      such branes. Applications include branes in $\Z_2$ orbifolds of 
      the SU(2) WZW model and in the D-series of unitary minimal models 
      considered previously by Runkel. 
      \noindent  
      \end{abstract}
      \vspace*{-20.9cm}
      {\tt {hep-th/0108126  \hfill  AEI 2001-102}} \break
      \bigskip\vfill
      \noindent\phantom{wwwx}{\small e-mail: }{\small\tt 
       keizo.matsubara@teorfys.uu.se, vschomer@aei.mpg.de,}\\  
        \phantom{wwwx {\small e-mail: }}{\small\tt 
        mikael.smedback@teorfys.uu.se} 
      \end{titlepage}

      \section{Introduction} 
      The study of branes at orbifold singularities has a long history. Such 
      investigations were motivated mainly by the fact that orbifolds of the 
      form $\QC^n/\Gamma$ can be used to locally model singularities in 
      Calabi-Yau spaces. Beyond such examples, more general orbifold 
      constructions are an important ingredient in conformal field theory 
      (CFT) model building, i.e.\ in the construction of exactly solvable 
      closed and open string backgrounds. In particular, Gepner's construction 
      of string theories on Calabi-Yau spaces involves some orbifold-like
      projection. This suggests to analyse strings and branes in orbifolds
      more general than $\QC^n/\Gamma$ and with orbifold actions that may 
      not admit an interpretation as a geometric symmetry of the target 
      space. 
      \medskip     
       
      It is not possible to give a complete account here of all 
      the previous results related to branes in orbifolds. Much  
      of the work was devoted to orbifold constructions in flat 
      space (see e.g.\ 
      \cite{DouMoo,AffOsh,Dou,DouFio,DiaGom,Gom,GabSte,Gab,CraGab,BiCrRo}). 
      The basis for most of these developments were laid in 
      \cite{DouMoo} which uses earlier ideas originating from 
      \cite{PraSag,GimPol}. Open string theory in more general conformal 
      field theory orbifolds was pioneered by Sagnotti and 
      collaborators starting from \cite{PraSag} (see also e.g.\  
      \cite{PrSaSt1,PrSaSt2}). Important contributions were made 
      later by Behrend et al. \cite{behrend,BePePeZu} and by Fuchs 
      et al. \cite{FucSch,FucSchI,BiFuSc}. The latter 
      extends the simple current techniques that were developed 
      for closed strings in \cite{SchYan1,SchYan2} to the case of
      open strings (see also \cite{HuScSo,Sche}). Work on open 
      strings in Gepner models  highlighting the orbifold aspects 
      includes \cite{GoJaSa,BruSch1,FuScWa,FKLLSW,FHSSW}. 

      All the contributions we have listed so far focus on the couplings 
      of open strings to the branes which is closely related to the 
      spectra of open strings that can stretch between branes in 
      orbifold spaces. The {\em operator product expansions} of open 
      string vertex operators (boundary fields) in an orbifold of 
      a non-trivial boundary conformal field theory, however,  were 
      addressed first in \cite{Run2} for minimal models and then later in 
      \cite{BruSch2} for A-type branes in Gepner models. We note 
      that such results on the couplings of open strings are a 
      necessary prerequisite if one wants to extend the studies
      of brane effective actions in orbifolds \cite{DouMoo} to 
      general backgrounds. 
      \medskip

      In this work we address the issue of boundary operator 
      product expansions for so-called {\em simple current 
      orbifolds} of the extension type. As was observed in 
      \cite{BruSch1,BruSch2}, their treatment follows very 
      closely the strategies known from orbifolds of a flat 
      background, i.e.\ they involve lifting the theory from the 
      orbifold to the covering space. The latter is described 
      by conformal field theory constructions that go back 
      to Cardy \cite{Car} and we will review them 
      along with some background material on simple current 
      symmetries in Section 2. After this preparation we enter 
      the central section of this work which contains our 
      main results (\ref{oope},\ref{resope}) on boundary operator 
      product expansions. Section 4 is devoted to applications. 
      We start by discussing a $\QZ_2$ orbifold of the SU(2) 
      WZW model. In this case, our algebraic results can be 
      interpreted geometrically. Finally, we reconsider the
      example of the D$_{even}$-series of unitary minimal 
      models which is contained in \cite{Run2} and we show 
      that our formulas provide a very elegant construction 
      of the solution.     

      \section{Boundary CFT on the covering space} 
      \setcounter{equation}{0} 

      As we mentioned in the introduction, our strategy for 
      dealing with branes on orbifold spaces follows the usual
      procedure in which the whole theory is lifted to a covering 
      space. We shall assume that the latter can be solved using 
      the standard microscopic techniques  which go back to the 
      work of Cardy \cite{Car}. The aim of this section is to 
      provide a brief account on this theory. 
      \medskip
       
      Suppose we are given some rational bulk conformal field theory 
      with chiral algebra $\cA = \bar \cA$ and a modular invariant 
      partition function of the form 
      \beq 
      Z(q,\bar q) \ = \ \sum_j \ 
      \chi_j(q) \ \chi_{\bj} (\bar q)\ \ \ . \label{Bpart}
      \end{equation} 
      Here $j$ runs through the sectors of the right moving chiral 
      algebra $\cA$ and each of these sectors $j$ comes paired with 
      a unique sector $\bj$ of the left moving chiral algebra $\bar 
      \cA = \cA$. As usual, $\chi_j(q)$ denotes the character of the
      sector $j$. 
      \smallskip

      The construction 
      of boundary theories involves picking some automorphism 
      $\Omega$ of the chiral algebra. This appears in the 
      boundary conditions to describe how left- and right 
      movers are glued along the boundary. Any such automorphism
      $\Omega$ induces a map $\omega$ acting on sectors $j$ 
      of the chiral algebra. Cardy's analysis of boundary 
      conditions applies whenever the partition function 
      (\ref{Bpart}) is ($\Omega$)-diagonal in the sense that 
      $\omega(j)^\vee = \bj$. Here, $j^\vee$ denotes the sector 
      conjugate to $j$, i.e.\ the unique label with the property 
      that its fusion product with $j$ contains the vacuum 
      representation $0$ of the chiral algebra. We shall assume 
      that $\Omega$ is chosen such that the modular invariant 
      partition function (\ref{Bpart}) of the covering theory 
      is $\Omega$-diagonal.%
      \smallskip%

      Under this condition, Cardy provides us with a list of 
      boundary theories. Their number agrees with the number 
      of sectors of $\cA$. We will use labels $I,J,K, \dots$ 
      to distinguish between boundary conditions and sectors 
      but it should be kept in mind that small and capital 
      letters run through the same index set. The spectrum 
      of open strings that stretch between the branes which 
      are associated with the labels $I$ and $J$ is given by 
      \beq Z_{IJ}(q) \ = \ \sum_j N_{Ij}^{\ \ J} \ \chi_j(q) 
      \ \ . \label{bpart} \end{equation}
      Obviously, this tells us how the state space $\cH_{IJ}$ of 
      the boundary theory is built up from sectors of the 
      chiral algebra. For a much more detailed explanation 
      of these results the reader is referred to \cite{ReSh1}.
      \smallskip

      There is a version of the state-field  correspondence
      in boundary conformal field theory that assigns a boundary 
      field to each state in $\cH_{IJ}$. Hence we can read off 
      {}from (\ref{bpart}) that the boundary primary field $\psi_j$ 
      appears with multiplicity $ N_{Ij}^{\ \ J}$ in the boundary theory. 
      The operator product expansion for two such primary fields is 
      given by 
      \beq
      \psi^{LM}_i(x_1) \ \psi^{MN}_j(x_2) \ = \ 
      \sum_{k} \ (x_1- x_2)^{h_k-h_i-h_j}\  
        \psi^{LN}_k (x_2) \ \Fus{M}{k}{i}{j}{L}{N}\ + \dots \ \ 
      \label{bOPE}
      \end{equation}
      for $x_1 < x_2$. Here, $F$ stands for the fusing matrix of the chiral 
      algebra $\cA$. It is defined as a linear transformation that relates 
      two different orthonormal bases in the space of conformal blocks 
      (see \cite{MooSei}) and it can be visualized as shown in 
      Figure 1. For simplicity we shall assume that the fusion rules 
      obey $N_{Ij}^{\ \ J} \leq 1$ so that the vertices carry no 
      additional labels.   
      \begin{figure}[htb]
      \centering
      \epsfig{file=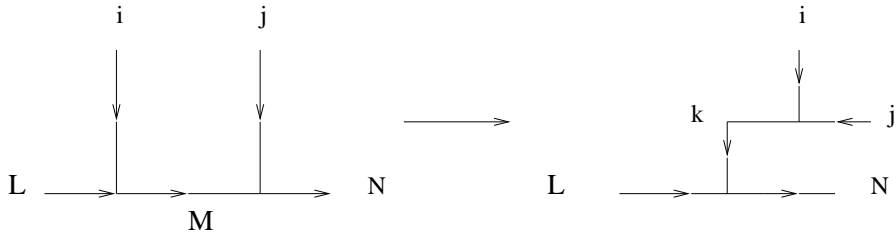, height=3cm}
      \caption{Graphical description of the fusing matrix}
      \label{fig:basic}
      \end{figure}

      The formula (\ref{bOPE}) was originally found for minimal 
      models by Runkel \cite{Run1} and extended to more general 
      cases in \cite{AlReSc1,FFFS1,FFFS2,behrend}. Note that our 
      conventions for the fusing matrix (see Figure 1) differ 
      slightly from the ones used in e.g.\ \cite{FFFS1,FFFS2}. 
      With the external legs being oriented as shown in Figure 1, 
      non of the six labels in the fusing matrix needs to be 
      conjugated.   

      In string theory, boundary operator products describe the scattering 
      of two open strings which are stretched between the branes $L,M$ 
      and $M,N$, respectively, into an open string that stretches 
      between $L$ and $N$. 

      Note that for the relation between the coefficients of the boundary 
      OPE and the fusing matrix it is crucial that boundary conditions 
      and boundary fields are labeled with elements from the same set. 
      This is no longer true for models with a `non-diagonal' (in the 
      sense specified above) bulk modular invariant partition function.  
      We shall see below how this can affect the boundary operator 
      product expansions. Examples of boundary OPEs for non-diagonal 
      modular invariants were studied in \cite{Run2,BruSch2} and they 
      are the main subject of this work.    
      \bigskip

      The formulas (\ref{bpart},\ref{bOPE}) provide a complete solution 
      of the open string sector on the covering space (for branes of 
      gluing-type $\Omega$). Before moving on to the orbifold theory 
      let us briefly study the symmetry properties of the solution 
      with respect to the group action that we plan to divide 
      out. We shall assume that this orbifold group is generated by 
      {\sl simple currents} of the conformal field theory. To describe 
      this more precisely, we need some new notation. Primaries (or the 
      associated conformal families) of a conformal field theory form a 
      set $\cJ$. Within this set $\cJ$ there can be non-trivial elements 
      $g \in \cJ$ such that the fusion product of $g$ with any other $j 
      \in \cJ$ gives again a single primary $g \times j = gj \in \cJ$. 
      Such elements $g$ are called {\em simple currents} and the set 
      ${\cal C}$ of all these simple currents forms an abelian subgroup 
      ${\cal C} \subset \cJ$. The product in ${\cal C}$ is inherited from 
      the fusion product of representations. From now on, let us fix some 
      subgroup $\Gamma \subset {\cal C}$. 
      \smallskip

      Through the fusion of representations, the index set $\cJ$
      comes equipped with an action $\Gamma \times \cJ \rightarrow
      \cJ$ of the group $\Gamma$ on labels $j \in \cJ$. Under this
      action, $\cJ$ may be decomposed into orbits. The space of these
      orbits will be denoted by $\cJ / \Gamma$ and we use the symbol
      $[j]$ to denote the orbit represented by $j \in \cJ$. These 
      orbits may have {\em fixed points}, i.e.\ there can be labels 
      $j \in \cJ$ for which the following stabilizer subgroup $\S_j 
      \subset \Gamma$
      \beq\label{stab}
      \S_{j} \ = \ \{ \ g \in \Gamma \ \mid\  g \cdot j \ = \ j\ \}
      \end{equation}
      is nontrivial. Up to isomorphism, the stabilizer subgroups
      depend only on the orbits $[j]$ not on the choice of a 
      particular representative $j \in [j]$, i.e.\ $\S_j = \S_{[j]}$.  
      \smallskip

      The last object we have to introduce is the  {\em charge} $\tQ_g(j)$ 
      of a primary $j$ with respect to the simple current $g$. It is 
      obtained from the following special matrix elements 
      $$\Omega \vert{j}{k}{i} \ = \ \Br{j}{i}{j}{i}{0}{k} $$  
      of the braiding matrix $B = B^{(+)}$ (see \cite{MooSei} for 
      details) by specialization to simple currents $i = g$, 
      \beq  
      (-1)^{\tQ_g(j)} \ := \Omega \vert{j}{gj}{g} \ \ .
      \end{equation}
      Note that this specifies $\tQ_g(j)$ up to an even integer, 
      i.e. $\tQ_g(j) \in \QR/ 2\QZ$. The charge $\tQ_g(j) $ is 
      related  to the more standard {\em monodromy charge}  
      $Q_g(j) \in \QR/\QZ$ by the prescription  $Q_g(j) := 
      \tQ_g(j)\ \mod\ 1$. We note that the monodromy charge
      can be computed from the conformal dimensions of the 
      involved fields through the expression  
      $$ Q_g(j) \ = \ h_j + h_{g} - h_{gj} \ \ \mod \ \ 1 \ \ . $$
      In case the simple currents have integer conformal weight
      (but not only then, see Section 4.2 for an example), the 
      monodromy charge $Q_g(j)$ depends only on the equivalence class 
      $[j]$ of $j \in J$. An orbit $[j]$ is said to be invariant, if 
      $Q_g([j]) = Q_g(j) = 0$ for all $g\in \Gamma$.
      \smallskip

      We are finally in a position to formulate the symmetry properties
      of the open string sector for the theory on the covering space. To 
      this end we introduce the following action of the simple current 
      group $\Gamma$ on boundary fields 
      \beq g(\, \psi^{IJ}_i(x) \, ) \ := \ (-1)^{-\tQ_g(i)} \, 
          \psi^{gI \, gJ}_i (x) \ \ . 
      \label{scact} 
      \end{equation}
      Here we use that in the Cardy theory the boundary labels $I,J$ are 
      taken from the set $\cJ$ which comes equipped with the action of 
      $\Gamma$ that we described above. Using the following symmetry 
      property of the fusing matrix \cite{BruSch2}    
      \beq
      \Fus{g M}{k}{i}{j}{gL}{g N} \ = \ 
       \Fus{M}{k}{i}{j}{L}{N}
       (-1)^{\tQ_g(i) + \tQ_g(j) - \tQ_g(k)}  
      \label{Fshift} 
      \end{equation}
      one can show that the operator product expansions (\ref{bOPE}) 
      respect the action of the simple current group $\Gamma$ on 
      boundary fields. In this sense, $\Gamma$ describes a symmetry 
      of the theory on the covering space. 

      \section{Boundary CFT on the orbifold} 
      \setcounter{equation}{0} 

      Our goal now is to discuss D-branes on an orbifold of the 
      original conformal field theory. Geometrically, one would 
      like to understand these branes on the orbifold space through 
      D-branes on the covering space. In such an approach, a brane 
      on the orbifold gets represented by several pre-images on the 
      covering space which are mapped onto each other by the 
      action of the orbifold group. As discussed in 
      \cite{BruSch1,BruSch2}, there is a large class of cases in 
      which these geometric ideas carry over to the construction 
      of branes in exactly solvable conformal field theories. 
      \medskip

      To begin with, let us give a precise formulation of our main 
      assumption on the partition function $Z^{\orb}(q)$ of the bulk 
      theory that we want to study. We assume that there is an orbifold 
      group $\Gamma$ within the group of all simple currents such that 
      $Z^{\orb}$ is of the form (see e.g.\ \cite{SchYan2})
      \beq\label{intmod}
      Z^{\orb}(q,\bar q) \ = \ \sum_{[j], Q_\Gamma ([j]) = 0 } \ 
        |\, \S_{[j]}\, | \
         | \sum_{j' \in [j]}\ \chi_{j'}(q) \, |^2 \ \  .  
      \end{equation}
      Note that this partition function does not have the simple form 
      (\ref{Bpart}) so that Cardy's theory for the classification and 
      construction of D-branes does not apply directly.  
      \medskip   

      An orbifold theory with bulk partition function of the form 
      (\ref{intmod}) possesses consistent boundary theories which 
      are assigned to orbits $[I]$ of labels $I$ that parametrize 
      the boundary theories of the parent CFT. The open string 
      spectra associated with a pair of such branes on the orbifold 
      are given by 
      \beq \label{Zproj}  
      Z^{\orb}_{[I][J]}(q) \ = \  
      \sum_{g,k} \ N_{I \,\, k}^{\ \ gJ} \chi_{k}(q)\ \ .
      \end{equation}
      This agrees precisely with the prediction from the geometric 
      picture of branes on orbifolds. In fact, the $I,J$ can be 
      considered as geometric labels specifying the position of the
      brane on the covering space. To compute spectrum of two branes 
      $[I]$ and $[J]$ of the orbifold theory, we lift $[I]$ to one of 
      its pre-images $I$ on the covering space and  include all the 
      open strings that stretch between this fixed brane $I$ on the 
      cover and an arbitrary pre-image $gJ$ of the second brane $[J]$. 
      \smallskip

      It is important to notice that in many cases the boundary conditions 
      $[I]$ can be further resolved, i.e.\ there exists a larger set 
      of boundary theories such that $[I]$ can be written as a 
      superposition  of boundary theories with integer 
      coefficients. This happens whenever the stabilizer subgroup
      $\S_{[I]}$ is non-trivial. In the absence of discrete torsion, 
      the elementary branes resolving the boundary condition $[I]$ 
      are labeled by characters $a,b,c,\dots$ of $\S_{[I]}$. 
      \footnote{More general possibilities including discrete 
      torsion have been discussed in \cite{Dou,DouFio,Gom,DiaGom,Gab,CraGab}. 
      The extension to general conformal field theory backgrounds
      can be found in \cite{FHSSW}} Geometrically, this corresponds to 
      the fact that the Chan-Paton factors of branes at orbifold fixed 
      points can carry different representations of the stabilizer 
      subgroup.  

      To spell out the spectrum of open strings stretching between two 
      such resolved branes $[I]_a$ and $[J]_b$ we need some more 
      notation. Let $H = H_{[I][J]} \subset \Gamma$ denote the subgroup 
      $\S_{[I]} \cap \S_{[J]}$ of our symmetry group $\Gamma$. The
      characters $e^{[I]}_a: \S_{[I]} \rightarrow $ U(1) and $e^{[J]}_b: 
      \S_{[J]} \rightarrow $ U(1) restrict to the common subgroup $H$ so 
      that the following numbers are well defined 
      \beq d^{\, a\,\, k}_{\ \ b} \ : = \ \frac{1}{|H|} \ \sum_{h \in H} \ 
         e_a(h) \, (-1)^{\tQ_h(k)} \, e_b(h^{-1})    \ \ . 
      \label{d} 
      \end{equation}
      With the chosen normalization, $d$ takes values in the set 
      $\{0,1\}$. The partition functions for the resolved branes 
      are given by (see \cite{BePePeZu,FucSch,FucSchI,behrend}
      for formulas that deal with general backgrounds)   
      \beq 
        Z^{\orb}_{[I]_a[J]_b}(q) \ = \ 
      \frac{1}{|\S_{[I]} \cdot\S_{[J]}|} \  
       \sum_{g,k} \ N_{I \,\, k}^{\ \ gJ} \ d^{\, a\,\, k}_{\ \ b} 
       \ \chi_{k}(q)\ \ . 
      \label{resopf} 
      \end{equation}  
      The normalization ensures that the coefficients appearing 
      in front of the characters $\chi_k(q)$ are integer. To see 
      this one should note that 
      $$ N_{I \,\, k }^{\ \ gJ} \ \neq \ 0 \ \ \ \ \Rightarrow \ \ \ \ \  
         N_{I \,\, k }^{ g_1 g g_2^{-1} J} \ \neq \ 0 \ \ 
        \mbox{ for all } \ \ g_1 \in \S_{[I]} \ , \ 
         g_2 \in \S_{[J]} \ \ . 
      $$ 
      This means that every term in eq.\ (\ref{Zproj}) comes
      with multiplicity 
      $$ |\S_{[I]} \cdot \S_{[J]}| \ = \ |\S_{[I]}|\, |\S_{[J]}|\,  
          |H|^{-1}  
      $$ 
      where the order of the subgroup $H$ appears because $H$ is isomorphic 
      to the kernel of the multiplication map $(g_1,g_2) \mapsto g_1 g_2^{-1} 
      \in \Gamma$. Let us remark that the partition functions 
      (\ref{resopf}) of the resolved branes sum up to the partition 
      function (\ref{Zproj}) of the projected boundary states. This
      follows easily from the property $\sum_{a,b} d^{a\, k}_{\ b}
      = |\S_{[I]} \cdot \S_{[J]}|$ of the constants $d$. Let us 
      note that the expression (\ref{resopf}) follows the usual 
      intuition that was developed in the context of orbifolds
      of the form $\QC^n/\Gamma$ (see e.g.\ 
      \cite{DouMoo,DiaGom,BiCrRo}). In fact, the formula guarantees 
      that an open string mode that transforms according to the 
      representation $\tQ_h(k): h \mapsto \tQ_h(k)$ of the subgroup 
      $H$ appears in the spectrum of strings stretching between 
      the branes $[I]_a$ and $[J]_b$ if and only if the representation 
      $e_a$ of the first brane $[I]_a$ together with the representation 
      $\tQ(k)$ add up to the representation $e_b$ of the second brane 
      $[J]_b$ (`conservation of charges').   
      \bigskip
        
      Restricting at first to unresolved D-branes, we will now 
      give explicit expressions for the operator products of boundary
      fields. Before writing them down, let us have another look at 
      eq.\ (\ref{Zproj}) and observe that for fixed $I,J,k$ there can 
      be several group elements $g \in \Gamma$ such that $N_{I\, \, k}^{gJ} 
      \neq 0$. We denote the associated subspace of $\Gamma$ by 
      \begin{equation} \label{gijk} 
      \Gamma \vert{k}{I}{J} \ = \ \{ \, g \in \Gamma \, |\, 
        N^{gJ}_{I\, \, k} \neq 0 \ \}\ \ . 
      \end{equation}  
      While the size of $\Gamma \vert{k}{I}{J}$ depends only on $k$ 
      and the orbits $[I],[J]$,  the subsets $ \Gamma \vert{k}{I}{J}$ 
      are selected depending on the choice of representatives $I \in [I]$ 
      and $J \in [J]$. If we shift these representatives along their 
      orbits, the $\Gamma \vert{k}{I}{J}$ shift according to   
      $$ \Gamma \vert{k}{I}{gJ} \ = \ g^{-1}\, \Gamma \vert{k}{I}{J} \ \ 
         \mbox{ and }  \ \ 
         \Gamma  \vert{k}{gI}{J} \ = \ g  \, \Gamma \vert{k}{I}{J} \ \ . 
      $$ 
      The group elements $g\in \Gamma \vert{k}{I}{J}$ label fields in the 
      boundary theory that describes open strings stretching between the 
      unresolved branes $[I]$ and $[J]$. We claim that they possess the 
      following  operator product expansions,   
      \begin{equation} \label{oope} 
       \Psi^{[L][M]}_{i,g_1} \ \Psi^{[M][N]}_{j,g_2}  \ = \ 
         \sum_k \ \Psi^{[L][N]}_{k,g_{12}} 
         \ (-1)^{-\tQ_{g_1}(j)}\, \Fus{g_1 M}{k}{i}{j}{L}{g_{12} N} \ + 
         \ \dots  \  
      \end{equation}
      where we suppressed the obvious dependence on world-sheet coordinates. 
      $L,M,N$ are representatives of the orbits $[L],[M],[N],$ and the group 
      element $g_{12}$ on the right hand side is given by  
      $$  
       g_{12} \ = \ g_1 \, g_2 \ \in\  \Gamma \vert{k}{L}{N} \ = \ 
        \Gamma  \vert{i}{L}{M}\, \cdot \, \Gamma  \vert {j}{M}{N} \ \ . 
      $$  
      Consistency of the boundary operator product expansion (\ref{oope}) 
      requires that the coefficients on the right hand side satisfy 
      certain sewing constraints that were first formulated by Lewellen 
      \cite{Lew} (see also \cite{PrSaSt2,Run1,Run2}). For the case at hand, 
      these are checked in the Appendix A.2. 
      \smallskip 

      Let us pause here for a moment and add two comments which can 
      provide some insight into the formula (\ref{oope}).  First, one 
      should observe that the operator product expansions we propose 
      mimic some kind of crossed product construction. \footnote{The 
      relation between crossed products and orbifolds is well known
      in string theory (see e.g.\ \cite{BerLei} and references therein).} 
      The formal similarities become most obvious if we think of the fields 
      in the orbifold theory as a product of the form $\Psi^{[I][J]}_k 
      \cdot g$ where the notation  separates the dependence on the 
      element $g \in \Gamma\vert{k}{I}{J}$ from the field. In 
      multiplying such composite objects, we would begin by moving 
      the $g_1$ from the first field through the second field. The 
      factor $\exp(-\pi i \tQ_{g_1}(j))$ encodes  the non-trivial 
      transformation law (\ref{scact}) of the second field under the 
      action of $g_1$. Following rel.\ (\ref{bOPE}), multiplication of 
      the two fields gives rise to a fusing matrix on the right hand 
      side of eq.\ (\ref{oope}) and the resulting field comes with a 
      factor $g_{12} = g_1 g_2$. Even though all this discussion is 
      very symbolic, it captures nicely the basic ingredients of 
      the formula (\ref{oope}). We will make the connection with 
      the crossed product more precise when we discuss the first 
      example below.
      \smallskip 

      In the special case that all the involved stabilizer groups
      $\S$ are trivial, we can obtain the expansions (\ref{oope}) 
      of the orbifold theory directly from the relations (\ref{bOPE}) 
      on the covering space. In fact, under the assumption of 
      trivial stabilizers, the new boundary fields in the orbifold 
      theory can be constructed by averaging the boundary fields of 
      the theory on the cover with respect to the group action, 
      i.e.\ 
      $$ \Psi^{[I][J]}_{i,g } (x) \ :=\ \sum_{g' \in \Gamma} 
         \ g'(\, \psi^{I\, g J}_i (x)\, ) \ \ . 
      $$ 
      Here we have chosen representatives $I,J$ of the orbits 
      $[I],[J]$. It is then easy to recover eq.\ (\ref{oope}) from the 
      corresponding expansions (\ref{bOPE}) on the cover.   
      \medskip

      When some of the boundary labels have non-trivial stabilizer 
      groups, the boundary fields of the unresolved theory must be 
      resolved according to the formula (\ref{resopf}). As we shall
      show below, this is achieved by the expressions 
      \begin{equation}\label{resfld}
       \Psi_{k,g}^{[I]_a[J]_b} = \sum_{g_1 \in S_I} \sum_{g_2 \in S_J}
       \Psi_{k,g_1 g g_2^{-1}}^{[I][J]}(-1)^{-\hat{Q}_{g_1}(k)}
      e_a(g_1)e_b(g_2^{-1}) \ \ . 
      \end{equation} 
      Here, $g$ runs through the set $\Gamma\vert{k}{I}{J}$ as above. 
      Note, however, that the set $\Gamma \vert{k}{I}{J}$ carries an 
      action of the subgroup $\cS_I \cdot \cS_J$. When we shift a field 
      of the form (\ref{resfld}) by $g_I \in \cS_I$ and $g_J \in \cS_J$ it 
      behaves according to 
      \begin{equation} \label{linrel}  
      \Psi_{k,g_Igg_J^{-1}}^{[I]_a[J]_b}\ = \ e_a(g_I^{-1}) 
       e_b(g_J) \,  \Psi_{k,g}^{[I]_a[J]_b}\ \ . 
      \end{equation} 
      After taking these relations into account, the space of boundary 
      fields is labeled by elements in the coset space $\Gamma \vert{k}{I}{J} 
      / \cS_I \cdot \cS_J$. Furthermore, it is easy to see that the expression 
      (\ref{resfld}) vanishes if $d^{a\, k}_{\ b} = 0$. These two observations 
      together show that the space of fields is in agreement with the partition 
      functions (\ref{resopf}). 

      The expression for the boundary operator product expansions of 
      the boundary fields (\ref{resfld})  can now be calculated from the 
      one for the unresolved fields. The result is  
      \begin{equation}\label{resope} 
      \Psi^{[I]_a[J]_b}_{i,g_1} \, \Psi^{[J]_c[K]_d}_{j,g_2} \ 
      =\  \delta_{b,c} \sum_{g\in \S_J,k} \Psi^{[I]_a[K]_d}_{k,gg_1g_2} 
      \, e_b(g) \, (-1)^{-\tQ_{gg_1}(j)} \, \Fus{g_1J}{k}{i}{j}{I}{gg_1g_2K} 
       + \dots \ \ \ ,
      \end{equation}
      where we have again neglected to spell out the obvious coordinate 
      dependence. In Appendix A.3 it is shown that the coefficients on 
      the right hand side of the above expression satisfy the appropriate 
      sewing constraint. Our formulas (\ref{resopf},\ref{resope}) 
      provide a complete solution of the open string sector in the 
      orbifold theory. 

      \section{Applications to WZW- and minimal models} 
      \setcounter{equation}{0}

      In the final section we want to outline some simple examples 
      of orbifolds that are covered by the general analysis of the 
      previous sections. These include open strings on a $\Z_2$ 
      orbifold of SU(2) and the so-called D$_{even}$-series 
      of minimal models. In the former case, our results admit a nice 
      geometric interpretation which will be discussed at the end
      of the first subsection. 
        
      \subsection{The $\Z_2$ orbifold of the SU(2) WZW model} 
      \def\Fun{{\mbox{\it Fun\/}}}

      The first example that we are going to discuss is given by orbifolds in 
      the SU(2) WZW model at level $k= 4n$ (see e.g.\ \cite{cft}). Let us start 
      with the diagonal bulk partition function describing the theory 
      before orbifolding. In our convention, the $k+1$ different sectors
      of the model will be labeled by $l = 0,1,...,k$ and the bulk 
      partition function is given by 
      \begin{equation}
      Z(q,\bar q) \ =\ \sum_{l=0}^{k} \, |\ \chi_l(q) \ |^2  \ \ . 
      \end{equation}
      The fusion product of any two sectors $l_1,l_2$ can be computed using 
      the standard rule 
      \beq \label{su2fus} 
      [l_1] \times [l_2] \ = \ [|l_1 -l_2|] +  [|l_1 -l_2| + 2] +  \dots +  
         [\min(l_1+l_2,2k-l_1 - l_2)]\ \ .
      \end{equation} 
      From this we can read off that the simple currents are given 
      by $l = 0$ and $l = k$. These two sectors form the group $\Gamma = \Z_2 
      = \{0,k\}$ which we will use in the orbifold construction.  
      \smallskip

      To specialize the general formula (\ref{intmod}) to our example we 
      need some preparation. Let us note first that the charges $\hat{Q}$
      defined in (\ref{Qdef}) can be found from the usual expression  
      for the braiding matrix of the WZW  model (see e.g.\ \cite{MooSei}),
      which implies
      \begin{equation}
        \Omega \vert{i}{l}{j}\ =\ (-1)^{\frac12 (i+j+l)}\, 
        e^{\pi i(h_i +h_j - h_l)}\ \ . 
       \end{equation}
      Here $h_{l} = l(l+2) /(4k+8)$ 
      is the conformal dimension of the sector $l$ at level $k$. Using these
      formulas for our simple currents $l=0$ and $l=k$, we obtain
      \begin{equation}
      \hat{Q}_0(j) \ = \ 0 \ \ \ \ \ , \ \ \ \   \hat{Q}_k(j)\ = \ \frac{j}{2}. 
      \end{equation}
      Orbits of $\Gamma = \Z_2$ consists of pairs $\{l,k-l\}$ as long as $l 
      \neq k/2$ and the sector $l=k/2$ leads to an orbit of length $1$ with 
      stabilizer group $\S_{[k/2]} = \Z_2$. An important condition for the 
      general theory to apply is that the members of a $\Gamma$-orbit have
      the same monodromy charge $Q = \hat Q \  \mod \ 1$. For the orbits $[l] = 
      \{ l, k-l\}$ this amounts to $l/2 = (k-l)/2 \ \mod \ 1$ which holds true
      since $k=4n$ is even. Finally, the orbits $[l]$ have vanishing 
      monodromy charge $Q$ if $l = 2m$ and the orbit $[k/2]= [2n]$ of length 
      1 belongs to this set. All these observations are summarized in the 
      following formula for the partition function of the orbifold theory,  
      \begin{equation}
      Z^{\orb}(q,\bar q) \ =\ \sum_{l=0}^{2n-2} |\chi_l (q) + \chi_{4n-l}(q)|^2  
           + 2|\chi _{2n}(q) |^2 
      \end{equation} 
      where the summation is over even $l$ only. $Z^{\orb}$ is known as 
      the D$_{\even}$ modular invariant of the SU(2) WZW model. 
      \bigskip

      According to the general theory our orbifold model has branes labeled 
      by $L = [0],[1], \dots, [k/2 - 1]$ and two additional branes $[k/2]_
      {\pm}$ associated with the two characters $e_+$ and $e_-$ of the stabilizer 
      group $\S_{k/2} = \Z_2$. We will not discuss them in full detail here
      but rather restrict to the most interesting case which appears when 
      the open strings have both ends on branes sitting at the fixed point. 
      Before resolution, the partition function (\ref{Zproj}) for these 
      open strings reads
      \beq Z^{\orb}_{[k/2][k/2]}(q) \ = \ \sum_{l=0}^{k}  \ 2\, \chi_l(q) 
      \label{Zprojsu} \end{equation}
      where summations runs over even $l$. To split this $Z^{\orb}$ into 
      the partition functions (\ref{resopf}) for resolved branes $[k/2]_{
      \pm}$ we compute the associated symbol $d$. In a matrix notation 
      $d^l = (d^{a\, l}_{\ \ b})$ it is given by 
      $$ d^{4p} \ = \ \left( \begin{array}{ll} 1 & 0 \\ 0 & 1 \end{array}
          \right)  \ \ \ \mbox{ and } \ \ \ 
         d^{4p+2} \ = \ \left( \begin{array}{ll} 0 & 1 \\ 1 & 0 \end{array}
          \right) $$
      with $p = 0, \dots, n$.  When inserted into eq.\ (\ref{resopf}), the 
      resolved partition functions become  
      \begin{equation} 
        Z^{\orb}_{[k/2]_{\pm}[k/2]_{\pm}}(q)\ = \ \sum_{p=0}^n \ \chi_{4p}(q) 
           \ \ \ ,  \ \ \ 
          Z^{\orb}_{[k/2]_{\pm}[k/2]_{\mp}}(q)\ = \ \sum_{p=1}^n \ \chi_{4p-2}(q)
           \ \ . \label{resopfsu}
      \end{equation} 
      These expressions are well known from previous studies of boundary 
      conditions in the D$_{\even}$ theory (see e.g.\ \cite{behrend}). Our 
      formula (\ref{resope}) describes the operator products of boundary fields
      in this model and thereby completes the solution of the model. 
      \bigskip 

      It is quite instructive to relate the operator 
      product expansions we found to the geometry of branes at the fixed 
      point. Recall from \cite{AlSc} that branes on the SU(2) are localized 
      along the k+1 `integer' conjugacy classes of SU(2)$=S^3$. The latter 
      are 2-spheres centered around the origin $e \in$ SU(2) with the 
      $(k/2)^{th}$ 2-sphere wrapping the equator of $S^3$. Since the 
      non-trivial element of our orbifold group $\Gamma = \Z_2$ acts by 
      {\em reflection} $ g \rightarrow -g$ along the equator of $S^3$, 
      the $(k/2)^{th}$ brane is located along the fixed surface of the 
      group action, in agreement with our algebraic results above. 
      \smallskip

      The algebra $\Fun(S^2)$ of functions on the equatorial 2-sphere 
      $S^2 \subset S^3$ is spanned by spherical harmonics $Y^{l}_\sigma$ 
      where $|\sigma| \leq l/2$ and $l$ is an even integer (recall that 
      we re-scaled all spins by a factor 2). This algebra inherits an 
      involution $\vartheta$ from the reflection on $S^3$. It acts on 
      spherical harmonics as $\vartheta(Y^l_\sigma) = (-1)^{l/2} 
      Y^l_\sigma$. The involution $\vartheta$ and its square $\vartheta^2 
      = \id$ give rise to an action of $\Z_2$ on $\Fun(S^2)$. Following a 
      general construction, we can use this data to pass to the {\em 
      crossed product} $\Fun(S^2) \times_{\vartheta} \Z_2$. This amounts to 
      extending the algebra of functions on $S^2$ by one additional 
      element $\theta = \theta_k$ subject to the conditions $\theta^2 
      = 1$ and
      $$ \theta \ Y^l_\sigma \ = \ \vartheta(Y^l_\sigma)\ \theta \ = \ 
         (-1)^{l/2} \, Y^l_\sigma \ \theta\  \ . $$
      Hence the new algebra is spanned by $Y^l_\sigma \theta_0 = Y^l_\sigma$ 
      and $Y^l_\sigma \theta$, i.e.\ its basis contains two SU(2) multiplets 
      for each even integer $l$. All this is very similar to the structure of 
      the partition function (\ref{Zprojsu}). In fact, we see that the latter 
      contains each representation of the SU(2) current algebra with 
      multiplicity $2$. For finite $k$, however, labels $l > k$ do not 
      appear in the partition function. This truncation in the spectrum of 
      boundary fields is related to the quantization of the spherical 
      branes of the SU(2) WZW model \cite{AlReSc1}. One can show that, 
      whenever the level $k$ is finite, all spherical branes on $S^3$ 
      carry a non-vanishing $B$-field. \footnote{In the $k \rightarrow 
      \infty$ theory, the B-field can be non-zero, but it vanishes on 
      the equatorial 2-sphere.}  

      The similarity between the crossed product geometry and the 
      open string theory goes much further. For $k \rightarrow  \infty$
      we may identify the basis elements of the crossed products with 
      the boundary primary fields according to 
      $$ V[Y^l_\sigma \theta_g](x) \:= \Psi^{[k/2][k/2]}_{l,\sigma;g}\ \ 
      \mbox{for } \ \ g \in \Z_2 \ \ .$$ 
      Here, $\sigma$ labels different boundary fields associated with 
      the ground states of the sector $l$ of the WZW-model. Following the 
      arguments in \cite{AlReSc1}, it is easy to see that this identification 
      preserves the multiplication, i.e.\ one finds 
      $$ \Psi^{[k/2][k/2]}_{i,\sigma_1;g_1}(x_1) \, 
           \Psi^{[k/2][k/2]}_{j,\sigma_2;g_2}(x_2) \ 
          \stackrel{k \rightarrow \infty}{\rightarrow} \ 
          V[Y^l_{\sigma_1} \theta_{g_1} Y^l_{\sigma_2} \theta_{g_2}](x_2) + 
          \dots 
      $$ 
      In the spirit of \cite{Vol,AlReSc1}, this shows that the 
      geometry of the unresolved equatorial brane is described 
      by the crossed product $\Fun(S^2) \times_{\vartheta} \Z_2$.   
      \smallskip

      Passing on to the geometry of the resolved branes, it is rather easy to 
      see that the elements $Y^l_{\sigma} \cdot (1 \pm \theta), l = 4m,$ 
      generate two 
      sub-algebras of $\Fun(S^2) \times_{\vartheta} \Z_2$. 
      These elements correspond to 
      primary boundary fields (\ref{resfld}) for open strings which have both 
      ends on the same resolved brane $[k/2]_{\pm}$. The rest of the basis in 
      the crossed product, namely the elements $Y^l_{\sigma} \cdot 
      (1 \pm \theta), 
      l = 4m-2,$ are associated with open strings that stretch between two 
      different resolved branes.
      \medskip

      We conclude this section with a few remarks on the dynamics of  
      branes in SU(2)$/\QZ_2$. Let us remark first that for a spherical 
      brane with fixed label $L$ the computation of the effective action 
      carries over from \cite{AlReSc1}. This means that one still finds
      a linear combination of a Yang-Mills  and a Chern-Simons term on 
      a fuzzy $S^2$ to control the behavior of these branes in the limit 
      $k \rightarrow \infty$. In particular, a stack of $D0$ branes at
      the origin is unstable and it can expand into a spherical brane.
      Following the reasoning of \cite{AlSc2,FreSch}, one can show that 
      these (unresolved) branes contribute a term $\QZ_{k/2+1}$ to the
      charge group of the background. In addition, the resolved branes 
      at the equator carry one extra charge that can be measured through
      the coupling of the closed string states in the twisted sector when 
      the level $k$ becomes large. 

      \subsection{D-branes in the D-series of minimal models}

      In this section we will analyze D-branes in minimal models. Our aim is 
      to show that our general theory is applicable to branes in the 
      D$_{even}$-series of minimal models. Their boundary operator product 
      expansions were first analysed by Runkel in \cite{Run2}, but the 
      resulting expressions where rather complicated and difficult to 
      work with. For the case of the D$_{even}$-series, our formulas 
      represent a considerable simplification. 
      \smallskip 

      {\it Minimal models} $M(p,p')$ are labeled by two integers $p$,$p'$ 
      \cite{cft}. Their chiral algebra is generated by Virasoro fields with 
      central charge
      \beq
        c \ = \ 1 - 6 \frac{(p-p')^2}{pp'}\ \ . 
      \end{equation}
      The primary fields are labeled by pairs $(r,s)$  with $1 \leq r \leq p'-1$ 
      and $1 \leq s \leq p-1$. We compute their conformal weights through the 
      formula 
      \beq \label{conformalweights}
        h_{r,s} \ = \ \frac{(pr-p's)^2-(p-p')^2}{4pp'}\ \ .
      \end{equation}
      Let us note that each sector of the model is represented by two pairs 
      of the form $(r,s)$. More precisely, the two labels
      \beq
      \label{hinvariance}
        (r,s)  \leftrightarrow (p'-r,p-s) 
      \end{equation}
      are associated with the same sector. This is consistent with the formula 
      (\ref{conformalweights}) for the conformal weights and it motivates to 
      introduce the fundamental region $E(p,p')$ containing pairs $(r,s)$ 
      which satisfy $p's < pr$. By construction, $E(p,p')$ contains each
      sector exactly once. For unitarity it is needed that $p=p'+1$ and 
      from now on we always assume this to be the case.
      \smallskip

      \def\sss{\scriptscriptstyle} 
      Let us proceed by investigating which simple currents are present in 
      this model. This can be read off from the well known fusion rules 
      \beq \label{fusionrules}
      (r,s) \times (m,n) \ = \ \sum_{
      \begin{array}{c} {\sss k=1+|r-m|} \cr
      {\sss k+r+m=1\, mod\, 2} \end{array}}^{k_{max}} \sum_{
      \begin{array}{c}{\sss l=1+|s-n|} \cr {\sss l+s+n=1\, mod \, 2} 
      \end{array}}^{l_{max}}\ (k,l)
      \end{equation}
      where 
      \begin{equation}
      \begin{split}
      k_{max} & = \mbox{min} (r+m-1,2p'-(r+m+1)) \\
      l_{max} & = \mbox{min} (s+n-1,2p-(s+n+1))
      \end{split}.
      \end{equation}
      Using these formulas together with the identifications (\ref{hinvariance}), 
      it is rather easy to see that the simple currents are given by the labels 
      $(1,1)$ and $(p'-1,1) \cong (1,p-1)$. Obviously, these two simple currents 
      generate a group $\Gamma = \{ (1,1) ,\ (p'-1,1) \} \cong \Z_2$. This is the
      group that we use in the orbifold construction. The orbits $[(r,s)]$ contain 
      the sectors $(r,s)$ and $(p'-r,s) \cong (r,p-s) $. A non-trivial stabilizer 
      $\cS_{(r,s)} = \Z_2$ appears if $r = p'/2$ or $s = p/2$. 

      Next, we shall construct the charge $\hat Q_g(j)$ of a sector $j$ with
      respect to a simple current $g$. Using the definition of $\hat{Q}_g(j)$ 
      from (\ref{Qdef}), and the relation $\Omega \vert{j}{gj}{g}=
      \exp \pi i (h_j+h_g-h_{gj}) $, we can conclude that
      \beq \label{Qdef} 
        \hat{Q}_g(j) \ = \ h_j + h_g -h_{gj} \quad \mbox{mod 2}\ \ .
      \end{equation}
      Inserting the formula  (\ref{conformalweights}) for conformal weights
      and the labels for elements of the orbifold group $\Gamma = \{ (1,1) ,
      \ (p'-1,1) \}$, the expression for $\hat Q$ reduces to
      \beq \label{minmodQhat}
      \begin{split}
        \hat{Q}_{(1,1)}(r,s)    & \ = \ 0 \quad \mbox{mod 2}  \\
        \hat{Q}_{(p'-1,1)}(r,s) & \ = \ -rs+1+\frac{1}{2}(pr+p's-p-p') 
      \quad \mbox{mod 2}\ \ . 
      \end{split}
      \end{equation}
      The condition $Q_{\Gamma}[(r,s)]=0$ of vanishing monodromy charge
      is satisfied for odd $r$ if $p'$ is even and for odd $s$ if $p$ is
      even. 

      Let us now write down the relevant bulk partition functions. The 
      A-series bulk partition function is given by the diagonal modular 
      invariant
      \beq
        Z_{A_{p'-1},A_{p-1}}=\sum_{(r,s) \in E(p,p')}|\chi _{r,s}(q) |^2.
      \end{equation}
      From this we can obtain the following D$_{even}$-series partition 
      functions by orbifolding with the simple current group $\Gamma = 
      \QZ_2 = \{(1,1),(p'-1,1)\}$ (see e.g.\ \cite{cft}),   
      \begin{equation} \label{ZD1}
      p'\, =\, 2(2m+1):  \quad Z_{D_{p'/2+1},A_{p-1}}\ = \ \frac{1}{2}\sum_{
      \begin{array}{c}
      {\sss (r,s) \in E(p,p')}\cr {\scriptstyle r\quad \, odd \,} 
       \end{array}}|\chi _{r,s} (q) +
      \chi_{p'-r,s}(q) |^2,
      \end{equation}
      \begin{equation} \label{ZD2}
      p\, =\, 2(2m+1): \quad Z_{A_{p'-1},D_{p/2+1}}\ =\ \frac{1}{2}\sum_{
      \begin{array}{c}
      {\sss (r,s) \in E(p,p')}\cr {\scriptstyle s\quad odd } \end{array}}
       |\chi _{r,s}(q) + \chi_{r,p-s}(q)|^2,
      \end{equation}
      where $m$ is an integer. Note that the unitarity condition $p = p'+1$ 
      implies that either $p$ or $p'$ is even but the conditions we have 
      formulated above require that neither $p$ nor $p'$ is a multiple
      of $4$. The partition functions of the D$_{even}$-series are precisely 
      of the form (\ref{intmod}). This can be seen using the properties of 
      the simple currents, the field identification (\ref{hinvariance}) and 
      the discussion from the previous paragraph. To explain the pre-factors 
      $1/2$ we have a short look at the first partition function 
      (\ref{ZD1}). If the first entry $r$ in the label $j=(r,s)$ is odd 
      then so is $p'-r$ in $gj=(p'-r,s)$. Consequently, each orbit $[j]$ 
      with trivial stabilizer will contribute twice to the sum over 
      $E(p,p')$. When $j=(r,s)$ has a non-trivial stabilizer subgroup, 
      the argument is slightly modified: From (\ref{intmod}),
      we infer that such a term will appear with a factor of 2. This is in 
      agreement with (\ref{ZD1}), since $\chi_{r,s}+\chi_{p'-r,s}=
      2\chi_{r,s}$ so that the character of a fixed point appears with 
      a factor $4$ before we multiply the whole partition function by 
      $1/2$. The same arguments apply to eq.\ (\ref{ZD2}). 
      \smallskip

      As discussed previously, an orbifold theory with a bulk partition 
      function of the type (\ref{intmod}) possesses consistent boundary 
      theories with partition functions given by formula (\ref{resopf}), 
      \beq
        Z^{\orb}_{[I]_a[J]_b}(q) \ = \sum_{k \in E(p,p')} 
          n_{[I]_a \ \ k}^{\ \ \ \ [J]_b}\   \chi_{k}(q)\ \ , 
      \end{equation}  
      where
      \beq
        n_{[I]_a \ \ k}^{\ \ \ \ [J]_b}\  = \ \frac{1}{|\S_{[I]} \cdot\S_{[J]}|}
          \ d^{\, a\,\, k}_{\ \ b} \ \sum_{g \in \Gamma} N_{I \,\, k}^{\ \ gJ} \ \ .
      \end{equation}
      Since $|\S_{[I]} \cdot \S_{[J]}| \ = \ |\S_{[I]}|\, |\S_{[J]}|\, |H|^{-1}$, 
      we conclude that $|\S_{[I]} \cdot \S_{[J]}| \ = 1$ if neither $[I]$ nor 
      $[J]$ has a non-trivial stabilizer subgroup. If at least one of them does, 
      the correct factor is instead $|\S_{[I]} \cdot \S_{[J]}| \ = 2$. Moreover, 
      we have $d^{(r,s)} = 1$ if either $[I]$ or $[J]$ has trivial stabilizer. 
      Otherwise, the labels $a,b$ can assume two different values  $a,b = \pm$ 
      and using the same matrix notations as in the previous subsection the 
      symbol $d$ reads 
      \begin{eqnarray} 
      d^{(r,s)}  & = & \left( \begin{array}{ll} 1 & 0 \\ 0 & 1
      \end{array}
          \right)  \ \ \ \mbox{ if } \ \ \ s \, =\,  1  \ \mod \ 4  \nn \\[2mm]
      d^{(r,s)} & = & \left( \begin{array}{ll} 0 & 1 \\ 1 & 0
      \end{array}
          \right) \ \ \ \mbox{ if } \ \ \  s \, = \, 3 \ \mod \ 4 \ \ . \nn 
      \end{eqnarray} 
      Here, we assumed that $p=2(2m+1)$. The same expressions for $d$ are 
      used in case of $p' = 2(2m+1)$ but with conditions depending on $r$ 
      rather than $s$. Collecting the results for this discussion, we 
      conclude that   
      \beq
      \begin{split}
        & (1) \ \ \ n_{[I]_a \ \ k}^{\ \ \ \ [J]_b}\ =\ N_{I \,\, k}^{\ \ J} +
      N_{I
      \,\, k}^{\ \ gJ} \\[1mm]
        & (2) \ \ \ n_{[I]_a \ \ k}^{\ \ \ \ [J]_b}\ =\ N_{I \,\, k}^{\ \ J} 
           \ = \ N_{I \,\, k}^{\ \ gJ} \\[1mm]
        & (3) \ \ \ n_{[I]_a \ \ k}^{\ \ \ \ [J]_b}\ =\ N_{I \,\, k}^{\ \ J} \ = \
      d^{a\,  k}_b\ \ ,  
      \end{split}
      \end{equation}
      where (1) is valid when neither stabilizer subgroup is non-trivial, 
      (2) applies to the case when precisely one stabilizer subgroup is 
      non-trivial and (3) describes the case of both stabilizer subgroups 
      being non-trivial. We note that these spectra coincide with the 
      partition functions that were studied by Runkel \cite{Run2}. Hence, 
      our general theory is indeed applicable to the D$_{even}$-series of 
      minimal models for the modular invariants (\ref{ZD1}) and (\ref{ZD2}),
      resulting in rather simple formulas for the  boundary operator product 
      expansions of the associated boundary theories. Most importantly, this
      opens the way for studies of brane dynamics in the D$_{even}$-series
      of minimal models along the lines of \cite{ReRoSc}.

      \section{Conclusions and outlook} 
      \setcounter{equation}{0} 

      In this work we provided the boundary operator product expansions 
      (\ref{oope},\ref{resope}) for a large class of orbifolds and 
      illustrated the results in two non-trivial examples. Whereas 
      the $\QZ_2$ orbifold of the SU(2) WZW-model admits a nice 
      geometric interpretation, there is no obvious geometry underlying 
      the D$_{even}$-series of the minimal models. It is one of the 
      advantages of boundary conformal field theory techniques that 
      they do not require the existence of any geometric orbifold
      action. 
      \medskip

      The most important applications of our results are associated with 
      the investigation of B-type branes in Gepner models. As proposed in 
      \cite{BruSch1}, these branes should be analysed as A-type branes on
      the mirror where the latter is realized as an orbifold using the 
      Greene-Plesser construction \cite{GrePle}. Together with the 
      projection on integer charges, the Greene-Plesser orbifold group 
      fits into the framework we have discussed here. To be precise, 
      let us note that discrete torsion can occur for B-type branes in 
      Gepner models \cite{FKLLSW,FHSSW}. Based on the formula (\ref{oope}) 
      for the unresolved branes, however, it is not difficult to include 
      such cases into our formalism. 
      \smallskip

      The boundary operator product expansions of the form (\ref{oope},
      \ref{resope}) may be used to calculate correlation functions of 
      open string vertex operators for a very large class of branes in 
      Gepner models. Computations of this type were initiated in 
      \cite{BruSch2} in the case of A-type branes and they can provide 
      important insight into the super-potential of branes deep in the 
      stringy regime. For B-type branes, this is particularly interesting 
      since one expects that such data do not require corrections 
      upon variation of the K\"ahler moduli. Hence, the conformal 
      field theory data may be compared with geometric results in the 
      large volume whenever the latter are available. Otherwise, conformal 
      field theory predicts e.g.\ the dimension of moduli spaces for 
      super-symmetric cycles. These issues are currently under 
      investigation.     
      \bigskip
      \bigskip

      \noindent
      {\bf Acknowledgments:} We wish to thank A.Yu.\ Alekseev, I.\ Brunner, 
      A.\ Recknagel, I.\ Runkel and E.\ Scheidegger for support and 
      discussions. K.M.\ and M.S.\ are grateful to INTAS for the grant 
      INTAS 99-01705, to the Swedish Institute for the grant SI/DAAD 
      7145/1997 and to the Max-Planck-Institut in Potsdam.  These three 
      organisations  supported a one month stay at the AEI Potsdam where 
      this work was initiated.  

      \begin{appendix} 
      \section{Appendix: Solution of the sewing constraints} 
      \setcounter{equation}{0} 

      Sewing constraints for the coefficients of operator product expansions 
      are consistency conditions which are obtained by taking different limits 
      in correlation functions \cite{Lew, Run1, Run2}.  In this section we wish
      to show that the solutions we propose do indeed satisfy the appropriate 
      sewing constraints. We begin by making some preliminary considerations. 
      Then we proceed to the cases at hand, the unresolved and the resolved 
      branes in the orbifold theory. 

      \subsection{Collection of relevant formulas}

      Let us begin by collecting a few relations which will prove to be useful 
      below. First, we note that the charges $\hat Q_g(k)$ provide a 
      representation of the simple current orbifold group $\Gamma$ in 
      the sense that 
      \begin{equation}
      \label{Qhat}
        \hat{Q}_{g_1 g_2}(k) \ = \ \hat{Q}_{g_1}(k)+\hat{Q}_{g_2}(k) \ \ , \ \ 
        \hat{Q}_{id}(k) \ = \ 0\ \ . 
      \end{equation}
      The latter equation is a consequence of the former. Moreover, from the 
      work \cite{MooSei}, we know that the fusing matrix $F$ obeys the 
      pentagon relation,
      \begin{equation}
      \label{Pentagon}
        \sum_s
        \Fus{p_2}{s}{j}{k}{p_1}{b}
        \Fus{p_1}{l}{i}{s}{a}{b}
        \Fus{s}{r}{i}{j}{l}{k}
        \ =\ 
        \Fus{p_1}{r}{i}{j}{a}{p_2}
        \Fus{p_2}{l}{r}{k}{a}{b}\ \ ,
      \end{equation}
      along with the following relation, expressing a symmetry of the 
      fusing matrix $F$, 
      \begin{equation}
      \label{Fsymmetry}
        \Fus{p}{i}{j}{k}{n}{l}
        \Fus{n}{0}{i}{i}{l}{l}
        \ = \ 
        \Fus{n}{k}{i}{j}{l}{p}
        \Fus{p}{0}{k}{k}{l}{l}\ \ . 
      \end{equation}
      Recall that in our conventions, the label $0$ corresponds to the 
      vacuum representation. Completing the list of properties of the 
      fusing matrix, we finally note the relation
      \begin{equation} \label{Fmatrixzero}
           \Fus{p}{0}{i}{i^\vee}{j}{k} \ =\ 0 \quad \mbox{ if  } 
          j \neq k^{\vee}
      \end{equation}
      To simplify notations we will assume that $j = j^\vee$ 
      throughout this appendix. Proofs for the general case 
      follow the same strategy but are a bit more tedious. 
      \medskip

      Based on the relations (\ref{Qhat})-(\ref{Fmatrixzero}) we will 
      now derive an equation that will serve as departing point for
      the proof of the sewing constraints. Before stating it, we need 
      some more notation. Let us choose four elements 
      \begin{equation}  g_1 \in \Gamma \vert{i}{I}{J} \ \
        \mbox{, } 
        g_2 \in \Gamma \vert{j}{J}{K} \ \ 
        \mbox{, }  \ \ 
        g_3 \in \Gamma \vert{k}{K}{L} \ \
        \mbox{, }
        g_4 \in \Gamma  \vert{l}{L}{I} \ \  
      \label{gich} \end{equation} 
      and use the shorthand notation $g_{12}$ for the product $g_1 g_2$ 
      etc. Furthermore, we pick group elements $g_I \in \cS_I$, $g_J
           \in \cS_J$ etc.\ satisfying the relation
      \begin{equation} \label{idrelation}
        g_1 g_2 g_3 g_4 g_I g_J g_K g_L \ = \ \id\ \ . 
      \end{equation}
      After this preparation we want to show that the constants 
      \begin{equation} \label{Csymbols}
        C_{(i,g_1)  (j,g_2)  (k,g_{12}\tilde{g})}^{[I]_a[J]_b[K]_c}
        :=
        \Fus{g_1 J}{k}{i}{j}{I}{g_{12}\tilde{g}K}
        (-1)^{-\hat{Q}_{g_1 \tilde{g}}(j)}
        e_b(\tilde{g})
      \end{equation}
      obey the following system of equations, 
      \begin{equation} \label{sewingrelbasic}
      \begin{split}
        & C_{(j,g_2)  (k,g_3)  (q,g_{23}g_K)}^{[J]_b[K]_c[L]_d}
          C_{(i,g_1)  (q,g_{23}g_K) 
      (l,g_{123}g_K g_J)}^{[I]_a[J]_b[L]_d} 
          C_{(l,g_{123}g_K g_J)  (l,g_4) 
      (0,g_{123}g_4 g_K g_J g_L)}^{[I]_a[L]_d[I]_a} \\[2mm] 
        &=
          \sum_p
          C_{(i,g_1)  (j,g_2)  (p,g_{12}g_J)}^{[I]_a[J]_b[K]_c}
          C_{(k,g_3)  (l,g_4)  (p,g_{34}g_L)}^{[K]_c[L]_d[I]_a}
          C_{(p,g_{12}g_J)  (p,g_{34}g_L)  (0,g_{12}g_{34}g_J g_L
      g_K)}^{[I]_a[K]_c[I]_a}
          \Fus{p}{q}{j}{k}{i}{l}\ \ . 
      \end{split}
      \end{equation}
      To prove this relation we spell out its right-hand side (rhs) 
      using (\ref{Csymbols}) and the eq. (\ref{Qhat}),  
      \begin{equation} \label{RHS1}
      \begin{split}
       \mbox{rhs} \ = \  \sum_p
        & \Fus{g_1 J}{p}{i}{j}{I}{g_{12}g_J K}\, 
          \Fus{g_3 L}{p}{k}{l}{K}{g_{34}g_L I}\, 
          \Fus{g_{12}g_J K}{0}{p}{p}{I}{I}\, 
          \Fus{p}{q}{j}{k}{i}{l}  \\[1mm]
        & (-1)^{-\hat{Q}_{g_1}(j)-\hat{Q}_{g_3}(l)-
          \hat{Q}_{g_{12}}(p)-\hat{Q}_{g_J}(p)
         -\hat{Q}_{g_J}(j)-\hat{Q}_{g_L}(l)-\hat{Q}_{g_K}(p)   }\ \ . 
      \end{split}
      \end{equation}
      We now want to apply the pentagon relation. To this end, let us 
      rewrite the first and third fusing matrix in the previous expression 
      with the help of eqs.\ (\ref{Fsymmetry}) and (\ref{Fshift}) for 
      $g=g_{34}$, i.e.\ 
      \begin{eqnarray} 
        \Fus{g_1 J}{p}{i}{j}{I}{g_{12}g_J K}
          \Fus{g_{12}g_J K}{0}{p}{p}{I}{I}  & = &  
         \Fus{g_{12}g_J K}{i}{j}{p}{g_1 J}{I}
          \Fus{g_1 J}{0}{i}{i}{I}{I} \nn \\[2mm] 
        &  & \hspace*{-3cm} = \   \Fus{g_{1234}g_J
      K}{i}{j}{p}{g_{134}J}{g_{34}I}
          \Fus{g_1 J}{0}{i}{i}{I}{I}
          (-1)^{ \hat{Q}_{g_{34}}(i)-\hat{Q}_{g_{34}}(j)-\hat{Q}_{g_{34}}(p) }
      \ \ . \nn \end{eqnarray}
      As for the second fusing matrix of expression (\ref{RHS1}), we shift it 
      using (\ref{Fshift}) with $g=g_{IL}^{-1}$. Altogether, these changes bring 
      (\ref{RHS1}) to a suitable form such that the pentagon relation 
      (\ref{Pentagon}) can be inserted to obtain 
      \begin{equation}
      \begin{split}
       \mbox{rhs} \ = \  & 
         \Fus{g_{1234}g_J K}{q}{j}{k}{g_{134} J}{g_3 g_{IL}^{-1} L}
          \Fus{g_3 g_{IL}^{-1} L}{i}{q}{l}{g_{134}J}{g_{34} I}
          \Fus{g_1 J}{0}{i}{i}{I}{I}    \\[1mm] 
        & (-1)^{ \hat{Q}_{g_{IL}}(k)+\hat{Q}_{g_{IL}}(l)+
          \hat{Q}_{g_{34}}(i)-\hat{Q}_{g_{34}}(j)
           -\hat{Q}_{g_1}(j)-\hat{Q}_{g_3}(l)
         -\hat{Q}_{g_J}(j)-\hat{Q}_{g_L}(l) }\ \ . 
      \end{split}
      \end{equation}
      The final step consists of yet another rewriting of all the fusing matrices.
      We shift the first fusing matrix  using (\ref{Fshift}) with $g=g_2g_{IKL}$,
      while the second is shifted with $g=g_{34}^{-1}$. The second and third 
      fusing  matrix can then be rewritten using (\ref{Fsymmetry}). If we also 
      take advantage of (\ref{Qhat}), the right-hand side is finally turned 
      into
      \begin{equation}
      \begin{split}
      \mbox{rhs} \ = \   & \Fus{g_2 K}{q}{j}{k}{J}{g_{23}g_K L}
          \Fus{g_1 J}{l}{i}{q}{I}{g_{123} g_{JK} L}
          \Fus{g_{123} g_{JK} L}{0}{l}{l}{I}{I}  \\[1mm] 
        & (-1)^{-\hat{Q}_{g_2}(k)-\hat{Q}_{g_1}(q)-\hat{Q}_{g_{123}}(l)
         -\hat{Q}_{g_{JK}}(l) -\hat{Q}_{g_K}(k)-\hat{Q}_{g_J}(q)
         -\hat{Q}_{g_L}(l) } \ \ .
      \end{split}
      \end{equation}
      This is identical to the left-hand side of (\ref{sewingrelbasic}), as
      can be seen using (\ref{Csymbols}) and it therefore completes our 
      derivation of the equation (\ref{sewingrelbasic}).

      \subsection{Sewing constraints for the unresolved case}

      In equation (\ref{oope}), we claim that the operator product expansion
      in the unresolved theory is
      \begin{equation}
      \label{OPE}
        \Psi_{i,g_1}^{[L][M]}(x_1)
        \Psi_{j,g_2}^{[M][N]}(x_2)
        \ =\ 
        \sum_k
        (x_1-x_2)^{h_i+h_j-h_k}
        \Psi_{k,g_{12}}^{[L][N]}(x_2)
        C_{(i,g_1) (j,g_2) (k,g_{12})}^{[L][M][N]}
        + \dots 
      \end{equation}
      where the coefficients are given by
      \begin{equation}
      \label{suggsolutionunresolved}
        C_{(i,g_1) (j,g_2) (k,g_{12})}^{[L][M][N]}
        \ =\ 
        \Fus{g_1 M}{k}{i}{j}{L}{g_{12}N}
        (-1)^{- \hat{Q}_{g_1}(j)}\ .
      \end{equation}
      Here and in the following the elements $g_i$ are taken from 
      subsets of $\Gamma$ that we have specified in (\ref{gich}). 

      The coefficients (\ref{suggsolutionunresolved}) must satisfy the 
      following version of the sewing relations (see \cite{Lew, Run1, Run2}),  
      \begin{equation}
      \begin{split}
      \label{sewingrelunresolved}
        & C_{(j,g_2) (k,g_3) (q,g_{23})}^{[J][K][L]}
          C_{(i,g_1) (q,g_{23}) (l,g_{123})}^{[I][J][L]}
          C_{(l,g_{123}) (l,g_4) (0,g_{123}g_4)}^{[I][L][I]} \\[2mm] 
        & =
          \sum_p\ 
         C_{(i,g_1) (j,g_2) (p,g_{12})}^{[I][J][K]}
          C_{(k,g_3) (l,g_4) (p,g_{34})}^{[K][L][I]}
          C_{(p,g_{12}) (p,g_{34}) (0,g_{12}g_{34})}^{[I][K][I]}
          \Fus{p}{q}{j}{k}{i}{l}
      \end{split}
      \end{equation}
      If $g_{1234} \not \in \S_I$ then both sides of the equation vanish 
      because the third operator product coefficient on either side is 
      zero. Hence we can assume that $g_{1234} \in \S_I$. We then derive
      the relation (\ref{sewingrelunresolved}) from the master equation 
      (\ref{sewingrelbasic}) by setting $g_J=g_K=g_L=e$ and 
      $g_I=g^{-1}_{1234} \in \S_I$.  

      \subsection{Sewing constraints for the resolved case}

      We turn now to the resolved case. Our claim is that the operator 
      product expansions for resolved boundary fields are given by 
      \begin{equation}
      \label{OPEa}
        \Psi_{i,g_1}^{[I]_a[J]_b}(x_1) \, 
        \Psi_{j,g_2}^{[J]_b[K]_c}(x_2) 
        \ = \ 
        \sum_{k,\tilde{g} \in S_J}\, (x_1-x_2)^{h_i + h_j - h_k} 
        \Psi_{k,g_1 g_2 \tilde{g}}^{[I]_a[K]_c}(x_2) \,  
         C_{(i,g_1)  (j,g_2)  (k,g_{12}\tilde{g})}^{[I]_a[J]_b[K]_c}
         +\dots 
      \end{equation}
      where the coefficients are given by
      \begin{equation}
      \label{suggsolutionresolved}
        C_{(i,g_1)  (j,g_2)  (k,g_{12}\tilde{g})}^{[I]_a[J]_b[K]_c}
        =
        \Fus{g_1 J}{k}{i}{j}{I}{g_{12}\tilde{g}K}
        (-1)^{-\hat{Q}_{g_1 \tilde{g}}(j)}
        e_b(\tilde{g})\ \ .
      \end{equation}
          
      The associated sewing relation is obtained from evaluating products 
      of four different boundary fields \cite{Lew, Run1, Run2} in two 
      different ways. Comparison of the contributions from the 
      identity fields gives  
      \begin{equation}
      \begin{split}
      \label{sewingrelresolved}
          \sum_{\tilde{g},\tilde{g}',\tilde{g}''}
        & C_{(j,g_2)  (k,g_3)  (q,g_{23}\tilde{g}'')}^{[J]_b[K]_c[L]_d}
          C_{(i,g_1)  (q,g_{23}\tilde{g}'') 
      (l,g_{123}\tilde{g}''\tilde{g})}^{[I]_a[J]_b[L]_d} 
          C_{(l,g_{123}\tilde{g}''\tilde{g})  (l,g_4) 
      (0,g_{123}g_4\tilde{g}''\tilde{g}\tilde{g}')}^{[I]_a[L]_d[I]_a}
          \Psi^{[I]_a[I]_a}_{0,\tilde{g}\tilde{g}' \tilde{g}''} \ = \  \\[2mm] 
         \sum_{g,g',g''} \sum_p 
        & C_{(i,g_1)  (j,g_2)  (p,g_{12}g)}^{[I]_a[J]_b[K]_c}
          C_{(k,g_3)  (l,g_4)  (p,g_{34}g')}^{[K]_c[L]_d[I]_a}
          C_{(p,g_{12}g)  (p,g_{34}g') 
      (0,g_{12}g_{34}gg'g'')}^{[I]_a[K]_c[I]_a} 
        \Psi^{[I]_a[I]_a}_{0,gg'g''}
          \Fus{p}{q}{j}{k}{i}{l}
      \end{split}
      \end{equation}
      where $g,\tilde{g} \in \S_J,  g',\tilde{g}' \in \S_L $ and $ g'',\tilde{g}'' 
      \in \S_K$. In spelling out these equations, we have decided to keep the 
      identity field. This allows us to take care of the obvious linear 
      relations (\ref{linrel}) that exist between our boundary fields.   
      \smallskip

      To take advantage of our relation (\ref{sewingrelbasic}), we choose to
      compare terms on the left-hand side and right-hand side such that
      $g=\tilde{g}=g_J$, $g'=\tilde{g}'=g_L$ and $g''=\tilde{g}''=g_K$
      where we have indicated our choice of the group elements 
      $g_J, g_L,g_K$ at the same time. As in the unresolved case, the 
      sewing relation holds trivially for $g_{1234} g_{IJK} \not \in \S_I$. 
      Otherwise, we can choose $g_I\in \S_I$ with the help of equation 
      (\ref{idrelation}) and obtain the terms in our sewing relations
      (\ref{sewingrelresolved}) from the corresponding terms in  
      equation (\ref{sewingrelbasic}). 
      \end{appendix}

      \def\gaw{Gawedzki}
      \newcommand{\sbibitem}[1]{\bibitem{#1}}
      
      \end{document}